\documentclass[aps,12pt,tightenlines]{revtex4}
\usepackage[dvips]{color,graphicx}
\usepackage{bm}
\begin{document}
\newcommand{\boldsigma}{\mbox{\boldmath $\sigma$}}\newcommand{\bfxi}{\mbox{\boldmath $\xi$}}
\newcommand {\boldgamma}{\mbox{\boldmath$\gamma$}}
\newcommand{\boldtau}{\mbox{\boldmath $\tau$}}
\newcommand{\bnull}{\bbox{0}}\newcommand{\bftau}{\mbox{\boldmath $\tau$}}
\newcommand{\bfx}{{\bf x}}
\newcommand{\bS}{\bbox{S}}\newcommand{\bfy}{{\bf y}}
\newcommand{\bfP}{{\bf P}}
\def\poinc{Poincar\'{e} }
\newcommand{\eq}[1]{Eq.~(\ref{#1})}\newcommand{\eqs}[2]{Eqs.~(\ref{#1},\ref{#2})}
\newcommand{\ba}{\begin{eqnarray}}
\newcommand{\bkt}{\bm{k}_{\scriptscriptstyle T}}
\newcommand{\mbk}{\bm{k}}\newcommand{\ea}{\end{eqnarray}}\def\bfq {{\bf q}}
\newcommand{\bpt}{\bm{p}_{\scriptscriptstyle T}}
\newcommand{\xbj}{x_{\scriptscriptstyle B}}
\def\bfm {{\bf m}}
\newcommand{\st}{{\scriptscriptstyle T}}\def\bfs {{\bf s}}
\def\bfn {{\bf n}}
\def\bfqp {{\bf q}_\perp}
\def\bfK{{\bf K}}
\def\bfKp{{\bf K}_\perp}
\def\bfL{{\bf L}}
\def\bfk{{\bf k}}
\newcommand{\bx}{\bm{x}}\def\bfp{{\bf p}}  
\newcommand{\bfkap}{\mbox{\boldmath $\kappa$}} 
\def\bfr{{\bf r}} 
\def\bfy{{\bf y}} 
\def\bfx{{\bf x}} 
\def\bfS{{\bf S}}
\def\be{\begin{equation}}
 \def \ee{\end{equation}}
\def\bea{\begin{eqnarray}}
  \def\eea{\end{eqnarray}}
\newcommand{\eqn} {Eq.~(\ref )}
\newcommand{\bb}{\langle}
\newcommand{\kk}{\rangle}
\newcommand{\bk}[4]{\bb #1\,#2 \!\mid\! #3\,#4 \kk}
\newcommand{\kb}[4]{\mid\!#1\,#2 \!\mid}

\def\notp{{\not\! p}}
\def\notk{{\not\! k}}
\def\up{{\uparrow}}
\def\down{{\downarrow}}
\def\bfb{{\bf b}}

\setlength{\textheight}{8.60in}
\setlength{\textwidth}{6.6in}
\setlength{\topmargin}{-.40in}
\setlength{\oddsidemargin}{-.125in}
\tolerance=1000
\baselineskip=14pt plus 1pt minus 1pt

\def\poinc{Poincar\'{e} }
\def\bfq {{\bf q}}
\def\bfK{{\bf K}}
\def\bfL{{\bf L}}
\def\bfk{{\bf k}}
\def\bfp{{\bf p}}  
\def\be{\begin{equation}}
 \def \ee{\end{equation}}
\def\bea{\begin{eqnarray}}
  \def\eea{\end{eqnarray}}
\def\eqn {Eq.~(\ref )}

\newcommand{\kx}[2]{\mid\! #1\,#2 \kk}
\def\notp{{\not\! p}}
\def\notk{{\not\! k}}
\def\up{{\uparrow}}
\def\down{{\downarrow}}
\def\bfb{{\bf b}}

\vspace{1.0cm}
\vspace{.50cm}
\title{\begin{flushright}{\normalsize NT@UW-07-10}
\end{flushright} 
 Densities, Parton Distributions, 
and Measuring the   Non-Spherical Shape of the Nucleon}

\author{Gerald A. Miller}
\affiliation{ University of Washington
  Seattle, WA 98195-1560}

\sloppy

\begin{abstract}Spin-dependent quark densities, matrix elements
of specific 
 density operators in 
proton states of definite spin-polarization, indicate that the nucleon may harbor 
an infinite 
variety of non-spherical shapes. We show that these matrix elements are closely
related to specific transverse momentum dependent
parton distributions accessible in the angular dependence of
the semi-inclusive processes 
 $ep\rightarrow e\pi X$ and the Drell-Yan reaction
$pp\rightarrow l\bar{l}X$. New measurements or analyses would
 allow the direct exhibition of the non-spherical nature of the 
proton. 
\end{abstract}
\maketitle
\vskip0.5cm

Since the discovery that that
the spins of quarks and anti-quarks account for only about 33\% of the nucleon
spin \cite{oldspin}, \cite{newspin}, many 
 experiments have sought the origins of the remainder, which
 must be accounted for by effects of quark and gluon
angular momentum. The importance of orbital angular momentum is also demonstrated 
in exclusive reactions.
 Measurements  \cite {Jones:1999rz,Gayou:2001qd} 
find that  the ratio of the proton's electric and magnetic form
factor $G_E/G_M$, falls with increasing momentum transfer $Q^2$ for 
1$<Q^2<6 $ GeV$^2$.
 This striking behavior  
 indicates that the sum of the orbital angular momentum of the 
quarks in the proton is non-vanishing
\cite{ralston,Braun:2001tj,Miller:2002qb,Ji}.

One expects that the presence of significant orbital angular momentum
would lead to a non-spherical shape, if such can be defined by an
appropriate operator.
 We showed  \cite{Miller:2003sa},   using 
the  proton  model of
Ref.~\cite{Frank:1995pv},
  that the   rest-frame ground-state 
matrix elements of spin-dependent
density operators
reveal a host of non-spherical shapes. 
The use of the spin-dependent density operator  allows the detailed connection between orbital, spin and total
angular momentum to be revealed in quantum systems.
In the model of \cite{Frank:1995pv} the orbital angular momentum originates from the relativistic nature of the
quarks manifest by lower components of  Dirac spinors of  the
wave function, but there are many other potential 
sources. 

It is natural to ask if  the non-spherical nucleonic shapes can be measured. While 
matrix elements of the non-relativistic spin-density operator 
have been measured in condensed
matter systems \cite{Prokes} to  reveal highly non-spherical densities and 
the related the orbital angular momentum content  of
electron orbitals, finding the corresponding determination of the nucleon properties
has remained a challenge. 
Our purpose 
here is to  show that  
matrix elements of the spin-dependent density are closely related to specific
unintegrated  (transverse momentum dependent)
parton densities that could be obtained by measuring the angular dependence
of the $ep\rightarrow e\pi X$ reaction and 
of the Drell-Yan production cross  section in
$ pp$ collisions.

We begin by explaining how  
the  shapes of a nucleon are  exhibited 
by studying the rest-frame ground-state 
matrix elements of spin-dependent
density operators \cite{Miller:2003sa}.
The usual  density operator in non-relativistic
quantum mechanics is given by 
\bea \widehat{\rho}(\bfr)= \sum_i 
\delta(\bfr-\bfr_i),\label{rho}\eea
where $\bfr_i$ is the position operator 
of the $i$'th particle. 
 Matrix elements of this operator yield the  density of
a system. Suppose the particles also have spin 1/2. Then one can measure
the probability that particle is at a given position $\bfr$ and has a 
spin in an arbitrary, fixed direction specified by a unit vector
 $\bfn.$ The  spin projection operator
is $(1+\boldsigma\cdot\bfn)/2 $, so the  spin-dependent density operator
is 
\bea \widehat{\rho}(\bfr,\bfn)= \sum_i
\delta(\bfr-\bfr_i){1\over2}(1+\boldsigma_i\cdot\bfn).\label{sddr}\eea 
The 
 spin-dependent density allows 
the presence of the orbital angular momentum to be revealed in
 the shape of the computed density.  
To understand this, it is worthwhile to consider a simple example of a
 single charged particle
moving in a fixed rotationally invariant potential in an energy eigenstate
$|\Psi\rangle$ 
of quantum
numbers: $l=1,j=1/2$, polarized in the  direction $\widehat{\bfs}$
 and radial wave function $R(r)$. We find
\bea \rho(\bfr,\bfn)=\langle \Psi 
\left\vert \widehat{\rho}(\bfr,\bfn)\right\vert \Psi 
\rangle
={R^2(r)\over 2}\bb  
\widehat{\bfs}\vert1+2\boldsigma\cdot \hat{\bfr}\;{\bfn} \cdot \hat{\bfr} -\boldsigma\cdot
{\bfn}\vert \widehat{\bfs}\kk.\eea
Suppose  $\hat{\bfn}$
 is either parallel or anti-parallel to
the direction of the proton angular momentum defined by the vector 
$\hat\bfs$. The direction of the
 vector $\hat\bfs$ defines an axis (the ``z-axis''), and
 the direction of vectors can be represented in terms of this axis:
$\hat{\bfs}\cdot\hat{\bfr}=\cos\theta$. With this notation
$ \rho(\bfr,{\bfn}=\hat{\bfs})={R^2(r)}\cos^2\theta,\;
 \rho(\bfr,{\bfn}=-\hat{\bfs}  )={R^2(r)}\sin^2\theta$ and
 the non-spherical shape is exhibited. 
The average of these
two cases   is  
a spherical shape, as is the  average over the direction of $\hat\bfs$ 
or the average over the direction of $\bfn$.

The densities of \eq{rho} and \eq{sddr} can  be extended to include other 
operators.
Indeed, Ref.~\cite{Prokes} uses only the spin-dependent term appearing
in \eq{sddr}, and this is weighted by the electronic charge. For systems of
quarks, the densities could be weighted by the charge of the quarks, or be
concerned with a specific flavor.
One could also weight the spin-dependence by other operators.
 In particular, consider
\bea
\widehat{\rho}_{\rm REL}(\bfr,\bfn)\equiv \sum_i 
\delta(\bfr-\bfr_i){1\over2}(1+\gamma^0_i\boldsigma_i\cdot\bfn),\label{sddrrel}\eea 
where the relativistic aspects are emphasized by the appearance of the Dirac operator
$\gamma^0$, which becomes unity in the non-relativistic limit. We denote
the density operators  of \eq{rho},\eq{sddr} and \eq{sddrrel}, 
and any number of obvious extensions,  simply as  densities.

These  densities  are  defined
in terms of position, but to use QCD it is necessary  to 
define operators that give
the probability for a particle to have a given momentum, $\bfK$, 
 and a given direction
of spin, $\bfn$.
The field-theoretic version of the spin-dependent charge 
 density operator, \eq{sddr},
 is a generalization of the operator defined in Ref.~\cite {Miller:2003sa}:
\bea&&\widehat{\rho}(\bfK,\bfn)=\int {d^3\xi\over(2\pi)^3} e^{-i\bfK\cdot\bfxi}
\left. \bar{\psi}(0)
(\gamma^0+\boldgamma\cdot\bfn\gamma_5){\cal L}(0,\xi;\;{\rm path})\psi({\bfxi})\right|_{t=\xi^0=0}, 
\label{qft}\eea 
where $\psi$ is a quark field operator and flavor indices are omitted. 
The quark  field
operators are evaluated at equal time and accompanied by a path-ordered 
exponential link operator
$ {\cal L}((0,\xi;\;{\rm path})$ 
needed for color-gauge invariance. 
This introduces a path-dependence, which must be specified correctly
 to obtain a 
parton interpretation. 
We will use the choice of Ref.~\cite{Mulders:1995dh}; see below. 
The first quantized version of \eq{qft} (neglecting the gluonic aspects) 
is of the form of \eq{sddr} except that the factor $\delta(\bfr-\bfr_i) $ is replaced
by $\delta(\bfK-\bfK_i)$, where $\bfK_i$ is the momentum of the i'th quark.
It is worthwhile to define another density 
corresponding to $\widehat{\rho}_{\rm REL}$ of \eq{sddrrel}.
This is given by
\bea&&\widehat{\rho}_{\rm REL}(\bfK,\bfn)=\int {d^3\xi\over(2\pi)^3} 
e^{-i\bfK\cdot\bfxi}
\left. \bar{\psi}(0)\gamma^0
(1+\boldgamma\cdot\bfn\gamma_5){\cal L}(0,\xi;\;{\rm path})
\psi({\bfxi})\right|_{t=\xi^0=0}. 
\label{qftrel}\eea

The matrix 
element of a   density operator in
a nucleon state $\vert P,S\rangle$ of definite total angular momentum  defined
 by  four-vector $S^\mu$ and momentum $P$ is
\bea
\rho(\bfK,\bfn,\bfS)
\equiv\bb P,S\vert \widehat{\rho}_{\rm G}(\bfK,\bfn)\vert P,S\kk, \;\rho_{\rm REL}(\bfK,\bfn,\bfS)
\equiv\bb P,S\vert \widehat{\rho}_{\rm REL}(\bfK,\bfn)\vert P,S\kk.
\label{sdd}\eea
The most general shape of the proton in its {\it rest frame}, 
obtained if parity and rotational invariance are upheld is then \cite{Miller:2003sa}
\bea
&&\rho(\bfK,\bfn,\bfS)=A(\bfK^2)+B(\bfK^2)\bfn\cdot\hat\bfS+C
(\bfK^2)\left(\bfn\cdot\hat{\bfK}\;\hat\bfS\cdot\hat{\bfK} -{1\over3}\bfn\cdot\hat\bfS\right)\nonumber\\
&&\rho_{\rm REL}(\bfK,\bfn,\bfS)=A_{\rm REL}(\bfK^2)+B_{\rm REL}(\bfK^2)\bfn\cdot\hat\bfS+C_{\rm REL}
(\bfK^2)\left(\bfn\cdot\hat{\bfK}\;\hat\bfS\cdot\hat{\bfK} -{1\over3}\bfn\cdot\hat\bfS\right)
\label{genshape}
,\eea
with the last terms generating the non-spherical shape. Any 
wave function that yields a non-zero value of the coefficient $C(\bfK^2)$ 
or $C_{\rm REL}(\bfK^2)$
represents a system
of a non-spherical shape. If the relativistic constituent quark model 
of \cite{Frank:1995pv} is used, the principle difference between 
$\rho(\bfK,\bfn,\bfS)$ and $\rho_G(\bfK,\bfn,\bfS)$ would be  that $C_{\rm REL}=-C$.
This indicates that either $C_{\rm REL}$ or $C$ can be  used to infer information about
the possible shapes of the nucleon.
Measuring either $C(\bfK^2)$ or $C_{\rm REL}(\bfK^2)$,
 would require 
 controlling the   
three different vectors $\bfn,\bfS$ and $\bfK$ or their equivalent.

The densities of \eq{genshape}  
are difficult to measure because the system must be probed without momentum
transfer and the initial and final states are the same. This configuration
 also appears
in parton distributions,
both ordinary and transverse-momentum-dependent TMD. 
Parton density operators  depend on pairs of quark-field operators defined
at a fixed  light cone time  $\xi^+=\xi^3+\xi^0=0$ while
the  density operators of 
 \eq{qft} and  \eq{qftrel} are defined as an equal-time, $\xi^0=0$, correlation 
functions  and 
cannot be regarded as parton density operators.
However, we find 
a relation
between the two sets of operators by 
  integrating  Eqs.~(\ref{qft},\ref{qftrel})
over all values of $K_z$. This sets $\xi^3=0$,  so  the quark field operators
 of \eq{qft} or   \eq{qftrel} are now evaluated at $\xi^0=0$ and $\xi^3=0$ so $\xi^\pm=0$:
\bea\widehat{\rho}_{T}(\bfK_T,\bfn)\equiv\int_{-\infty}^\infty dK_z\widehat{\rho}(\bfK,\bfn) 
=\int {d^2\xi_T\over(2\pi)^2} e^{-i\bfK_T\cdot\bfxi_T}
\left.\bar{\psi}(0)
(\gamma^0+\boldgamma\cdot\bfn\gamma_5){\cal L}( 0,\xi;\;n^-)\psi({\bfxi_T})\right|_{\xi^\pm=0}
\nonumber\\
\widehat{\rho}_{{\rm REL}T}(\bfK_T,\bfn)\equiv\int_{-\infty}^\infty dK_z\widehat{\rho}_{\rm REL}(\bfK,\bfn) 
= \int {d^2\xi_T\over(2\pi)^2} e^{-i\bfK_T\cdot\bfxi_T}
\left.\bar{\psi}(0)\gamma^0
(1+\boldgamma\cdot\bfn\gamma_5){\cal L}( 0,\xi;\;n^-)\psi({\bfxi_T})\right|_{\xi^\pm=0}
\label{qftt}\eea 
where the specific 
path $n^-$ is  that of Appendix B of \cite{Mulders:1995dh}.
To obtain the relevant transverse  densities we take the matrix element of 
$\widehat{\rho}_{GT}$ in a nucleon state polarized in the transverse direction
$\bfS_T$, with 
\bea
&&\rho_{T}(\bfK_T,\bfn,\bfS_T)
\equiv\bb P,\bfS_T\vert \widehat{\rho}_{T}(\bfK_T,\bfn)\vert P,\bfS_T\kk\label{nrt}\\&&
= A_{T}(K_T^2) +B_{T}(K_T^2)\bfn\cdot\hat\bfS_T+
C_{T}(K_T^2)\frac{(\bfn\cdot\bfK_T \hat\bfS_T\cdot\bfK_T-{1\over2}K_T^2\bfn\cdot\hat\bfS_T)}{M^2},\nonumber\\&&
\rho_{{\rm REL}T}(\bfK_T,\bfn,\bfS_T)=A_{{\rm REL}T}(K_T^2) +B_{{\rm REL}T}(K_T^2)\bfn\cdot\hat\bfS_T+
C_{{\rm REL}T}(K_T^2)\frac{(\bfn\cdot\bfK_T 
\hat\bfS_T\cdot\bfK_T-{1\over2}K_T^2\bfn\cdot\hat\bfS_T)}{M^2},
\label{sddt}\eea
where $M$ is the nucleon mass, and we take the unit vector $\bfn$ 
to be in the transverse direction.

Information regarding the shape of the nucleon resides in the functions
$C_{T},C_{{\rm REL}T}$.
 We  now are able to  connect our newly-defined  transverse densities
with TMD parton distributions. The latter are related to 
Dirac projections of correlation functions \cite{Mulders:1995dh}:
\begin{eqnarray}
\Phi^{[\Gamma]}(x,\bfK_T) & = &
\left. \int \frac{d\xi^-d^2\xi_\st}{2\,(2\pi)^3} 
\ e^{iK\cdot \xi}
\,\langle P,S \vert \overline \psi (0)\,\Gamma\,{\cal L}(0,\xi;n_-)
\,\psi(\xi) \vert P,S \rangle \right|_{\xi^+ = 0}. \label{projection}
\end{eqnarray}
The projections $\Phi^{[\Gamma]}$ depend on the fractional momentum
$x$ = $K^+/P^+$, $\bfK_T$ and  on the hadron momentum
$P$ (in essence only $P^+$ and $M$ where we work in a frame in which $P^+\gg M$.).
Depending on the Lorentz structure of the Dirac matrix $\Gamma$ the
projections $\Phi^{[\Gamma]}$ are ordered according to powers
of $M/P^+$ multiplied with a function depending only on $x$ and $\bfK_T^2$. 
Each factor $M/P^+$ 
leads to a suppression by a power  in cross sections 
\cite{Levelt-Mulders-94a} so that   one  
may  refer to the projections
as having a 'twist' $t$ related to the power $(M/P^+)^{t-2}$ that
appears. Then moments (in $x$) of the $\bfK_T$-integrated 
functions  involve local operators of twist $t$ 
\cite{Collins-Soper-82}.

We use certain transverse momentum dependent parton distribution  functions of 
\cite{Mulders:1995dh}, which for transversely polarized nucleons are
given by 
\begin{eqnarray}
& & \Phi^{[\gamma^+]}(x,\bfK_T) =
f_1(x ,K_T^2) ,
\label{chirality}
\\ & & 
\Phi^{[ i \sigma^{i+} \gamma_5]}(x,\bfK_T) 
= S_T^i\,h_1(x ,K_T^2)
+ \frac{\left(K_T^i K_T^j 
- \frac{1}{2}K_T^2\delta_{ij}\right) S_T^j}{M^2}
\,h_{1T}^\perp(x ,K_T^2),
\label{transversity}
\end{eqnarray}
\begin{eqnarray}
 & & \Phi^{[ \gamma^i \gamma_5]}(x,\bfK_T) 
= \frac{M\,S_T^i}{P^+} \, 
g_T(x ,K_T^2)
+ {M\over P^+}\frac{\left(K_T^i K_T^j 
- \frac{1}{2}K_T^2\delta_{ij}\right) S_T^j}{M^2}
\,g_T^\perp(x ,K_T^2).\label{gt}
\end{eqnarray}
Terms of higher order in $(M/P^+)$ are neglected in the extraction of the 
functions $g_T,g_T^\perp$ from high energy data. 
Similarly, at high  energies,  we may replace $\gamma^+$ by $\sqrt{2}\gamma^0$.  
Note that the quantity $g_T^\perp$ is closely related to the
quantity $C_T$, while the quantity $h^\perp_{1T}$ is closely related to 
$C_{{\rm REL}T}$ 
because  $i\sigma^{i+}\gamma^5=
\gamma^+\gamma^i\gamma^5\rightarrow \sqrt{2} \gamma^0\gamma^i\gamma^5$.
Extracting $g_T^\perp$ would require a higher twist analysis, while 
$h^\perp_{1T}$ appears at leading order in the cross sections for semi-inclusive
leptoproduction experiments \cite{Boer:1997nt}. Thus the relativistic
spin-dependent density \eq{sddt} is easier to measure than the quantity of 
\eq{nrt}.

We integrate the above parton distribution functions 
over all $x$  so that the field operators are 
evaluated at $\xi^\pm=0$ as in  our  spin-dependent  
densities. A tilde is placed over a given quantity to define the $x$-integrated
result, {\it e.g.} 
$\widetilde{\Phi}^{[\Gamma]}(\bfK_T)\equiv \int\;dx \Phi^{[\Gamma]}(x,\bfK_T),\;
\widetilde{f}_1(K_T^2)\equiv\int\;dx f_1(x,K_T^2) $,
{\it etc.}
Then
\begin{eqnarray}
\widetilde\Phi^{[\Gamma]}(\bfK_T)= \left. \int \frac{d^2\xi_\st}{2P^+\,(2\pi)^2} 
\ e^{-i\bfK_T\cdot \bfxi_T}
\,\langle P,S \vert \overline \psi (0)\,\Gamma\,{\cal L}(0,\xi;n_-)
\,\psi(\bfxi_T) \vert P,S \rangle \right|_{\xi^+ = 0,\xi^-=0}.\label{projectiont}
\end{eqnarray}
The various $\Phi^{[\Gamma]}$ are expressed, in the infinite momentum frame,
in terms of transverse momentum dependent  
parton distribution functions \eq{transversity}
and \eq{gt}.
To relate these to our functions $A_T,B_T,C_T$ we evaluate these
equations in the rest frame ($P^+\rightarrow M/\sqrt{2}$, 
$\gamma^+\rightarrow\sqrt{2}\gamma^0$)  and choose the 
 operators $\Gamma$  
to correspond to those appearing in the spin-dependent 
densities.
 We find that 
\bea &&
\sqrt{2}{\rho}(\bfK_T,\bfn,\bfS_T)=\tilde{f}_{1}(K_T^2)+\tilde{g}_T(K_T^2)\bfn\cdot\hat\bfS_T
+{\left(\hat\bfn_T\cdot\bfK_T \hat\bfS_T\cdot\bfK_T-{1\over2}K_T^2\hat\bfn\cdot\hat\bfS_T\right)\over M^2}
\tilde{g}_{T}^\perp(K_T^2),\label{rhot}\\
&&\sqrt{2}{\rho}_{{\rm REL}T}(\bfK_T,\bfn,\bfS_T)=\tilde{f}_{1}(K_T^2)+\tilde{h}_{1}(K_T^2)\bfn\cdot\hat\bfS_T
+{(\hat\bfn_T\cdot\bfK_T \hat\bfS_T\cdot\bfK_T-{1\over2}K_T^2\hat\bfn\cdot\hat\bfS_T)\over M^2}
\tilde{h}^\perp_{1T}(K_T^2).\nonumber\\\label{rhotrel}
\eea
Finding 
a non-zero value  of either $\tilde{g}_T$ or $\tilde{h}^\perp_{1T}$
 would demonstrate that the proton is not spherical. 
Both  the  spectator
model \cite{Jakob:1997wg} and  the     quark model \cite{Frank:1995pv}
yield the result that  $\tilde{g}_T=-\tilde{h}^\perp_{1T}.$ 
This relation  does not appear to be a  general result.
In particular, one may use the most general parameterization of the correlation
functions \cite{Mulders:1995dh,Goeke:2005hb} in terms of scalar functions
$A_i(K,P)$ to show that
$g_T^\perp$ is proportional to the term $A_{8}$ and $h_{1T}^\perp$ is proportional
to the term $P^+/M A_{11}$. This means that, in the nucleon rest frame,
the ratio $g_T^\perp/h_{1T}^\perp$
is  simply a function of $(x,\bfK_T^2)$.  Furthermore, 
the  quantity $\tilde{h}^\perp_{1T}$ is known to characterize
 the dependence of the 
transverse
polarization of quarks in a transversely polarized nucleon
on the direction of $\bfK_T$ \cite{Boglione:1999pz,Barone:2001sp}.
Thus the  two quantities  $\tilde{g}_T^\perp$ and 
$\tilde{h}^\perp_{1T}$ each  characterize the non-spherical nature of
the nucleon, and the density $\rho_{{\rm REL}T}$ can be thought of
as ``the'' spin-dependent density''.

Next we  focus on experimental means to access $\tilde{h}^\perp_{1T}$.
The presence of the term 
$\tilde{h}^\perp_{1T}$ causes  distinctive signatures in 
semi-inclusive leptoproduction experiments  
 \cite{Boer:1997nt} in which  a hadron $h$ is produced. If the target is polarized in a  direction transverse to the 
lepton scattering plane, 
the cross
section acquires a term proportional to $\cos (3\phi_h^l)$ where
$\phi_h^l$  is the angle between the hadron production plane (defined by the momenta
of the incoming virtual photon and the outgoing  hadron) and the lepton scattering
plane. A similar effect occurs in electroweak semi-inclusive deep inelastic
leptoproduction and this  could be accessed  at high energies such as those found
at  HERA  \cite{Boer:1999uu}. Another signature occurs in the angular 
distribution of the  leptoproduction of
$\rho$ mesons \cite{Bacchetta:2000jk}, obtained 
 using an unpolarized lepton beam and a 
transversely polarized target. Similarly the term $\tilde{h}^\perp_{1T}$ makes its
presence felt in 
studying the production of two-pions inside 
the same current jet \cite{Radici:2001na}.
In each of these cases, the momentum of the virtual photon and its
vector nature provide the analogue of two of the three vectors $\bfn$ and $\bfS_T$ 
needed
to define the spin-dependent density. The hadronic  transverse momentum provides
the third, $\bfK_T$.
 
Another interesting possibility occurs 
in the Drell-Yan reaction $pp(\uparrow)\rightarrow l\bar{l} X$
 using one transversely polarized proton \cite{Boer:1999mm}.  
In case the  term ${h}^\perp_{1T}$  causes a distinctive oscillatory 
 dependence
on the angle $3\phi-\phi_{S_{1}},$ where 
$\phi$ is the angle between  the momentum of the outgoing lepton and the reaction 
plane in the lepton center of mass frame, and $\phi_{S_{1}}$ denotes the direction
of polarization with  respect to the reaction plane.  The  term ${h}^\perp_{1T}$ is 
multiplied by the anti-quark Boer-Mulders function.

To illustrate the shapes that could be obtainable using this method we
use the spectator model of \cite{Jakob:1997wg}
to evaluate the shapes of the proton. We rewrite \eq{rhotrel} as
\bea
{\sqrt{2}\rho_{{\rm REL}T}(\bfK_T,\bfn)\over\tilde{f}_1(K_T^2)}=1+
{\widetilde{h}_1(K_T^2)\over\tilde{f}_1(K_T^2)}\cos\phi_n+
{1\over2}{K_T^2\over M^2}
\cos(2\phi-\phi_n){\widetilde{h}_{1T}^\perp(K_T^2)\over\tilde{f}_1(K_T^2)},\label{shapeq}\eea
where $\phi$ is the angle between $\bfK_T$ and $\bfS_T$ and $\phi_n$ is the
angle between $\bfn$ and  $\bfS_T$. 
The transverse shapes of the nucleon, as defined by the right hand side of 
\eq{shapeq} are shown in Fig.~1, taking $\phi_n=0$.
Deformation is seen for values of $K_T$ as small as 0.25 GeV, and this increases
as $K_T$ increases.
Choosing  $\phi_n=\pi$ emphasizes the non-spherical nature
because the first two terms of \eq{shapeq} tend to cancel. 
The possible shapes implied by \eq{shapeq}
can be thought of as transverse projections of the shapes displayed in 
Refs.~\cite{Miller:2003sa}.
\begin{figure}
\includegraphics[width=4cm]{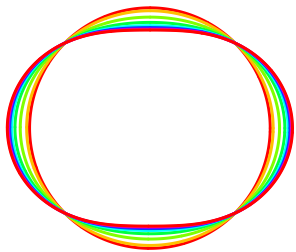}
\label{shape}
\caption{(Color online) Transverse shapes of the nucleon:
$\sqrt{2}\hat\rho_T(\bfK_T,\bfn)/\tilde{f}_1(K_T^2)$. The horizontal axis is the the direction of $\bfS_T$ and $\bfn=\hat{\bfS}_T,\; \phi_n=0$.
 The shapes vary from  circular to highly deformed
as $K_T$ is increased from 0 to 2.0 GeV in steps of 0.25 GeV. }
\end{figure}
\begin{figure}
\includegraphics[width=4cm]{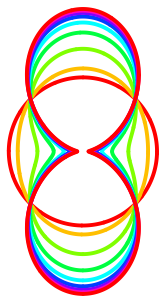}
\label{shape1}
\caption{(Color online) Transverse shapes of the nucleon, as in Fig.~1 except that
$\phi_n=\pi$.}
\end{figure}
One complication as that it would be very difficult to measure  the necessary
TMD's at all values of $x$ to construct the integrals appearing  here.
However, the model \cite{Jakob:1997wg} indicates that the
functions $f_1,h_1$ and $h_{1T}^\perp$ have very similar $x$ dependence, so that
measurements at values of $x$ for which these functions peak should be sufficient.

We have shown that
that the non-spherical nature of the nucleon shape is closely related to 
the non-vanishing
of the measurable TMD  $h_{1T}^\perp$. 
While determining this function experimentally represents a challenge, 
the ultimate determination of a non-zero values would 
clearly demonstrate that the shape of the proton not  round.

\section*{Acknowledgments}
We thank the USDOE for partial support of this work, D. Boer, 
W. Detmold and L. Gamberg
for useful
discussions and J.C. Peng for a stimulating oral presentation.

\end{document}